\documentclass{article}
\usepackage{spconf,amsmath,graphicx,hyperref}
\usepackage{algorithmic,algorithm}
\usepackage{amsmath,bm,graphicx}
\usepackage{cite}
\usepackage{url}
\usepackage{tabularx}
\usepackage{algorithmic,algorithm}
\usepackage{graphicx}
\usepackage{amsmath,bm,graphicx}
\usepackage{amsfonts}
\usepackage{multirow}
\usepackage{booktabs}
\usepackage{diagbox}
\usepackage{hyperref}

\title{Enhancing Noise Robustness for Neural Speech Codecs through Resource-Efficient Progressive Quantization Perturbation Simulation}
\name{Rui-Chen Zheng,~Yang~Ai,~Hui-Peng Du,~Li-Rong~Dai \thanks{This work was funded by the National Nature Science Foundation of China under Grant 62301521 and the Anhui Provincial Natural Science Foundation under Grant 2308085QF200.}}
\address{National Engineering Research Center of Speech and Language Information Processing, \\University of Science and Technology of China, Hefei, P. R. China\\
{\small \tt \{zhengruichen, redmist\}@mail.ustc.edu.cn,  \{yangai, lrdai\}@ustc.edu.cn}}
\begin{document}
\ninept
\maketitle
\begin{abstract}
Noise robustness remains a critical challenge for deploying neural speech codecs in real-world acoustic scenarios where background noise is often inevitable. A key observation we make is that even slight input noise perturbations can cause unintended shifts in quantized codewords, thereby degrading the quality of reconstructed speech. Motivated by this finding, we propose a novel and resource-efficient training strategy to enhance the noise robustness of speech codecs by simulating such perturbations directly at the quantization level. Our approach introduces two core mechanisms: 
(1) a distance-weighted probabilistic top-K sampling strategy that replaces the conventional deterministic nearest-neighbor selection in residual vector quantization (RVQ); and (2) a progressive training scheme that introduces perturbations from the last to the first quantizer in a controlled manner.
Crucially, our method is trained exclusively on clean speech, eliminating the need for any paired noisy-clean data.
Experiments on two advanced neural speech codecs, Encodec and WavTokenizer, demonstrate that the proposed strategy substantially improves robustness under noisy conditions—for example, boosting UTMOS from 3.475 to 3.586 at 15 dB SNR on Encodec—while also enhancing coding quality for clean speech.
\end{abstract}
\section{Introduction}
\label{sec:intro}

Speech codecs are essential tools that compress speech into discrete codes and reconstruct waveforms with minimal perceptual degradation. They play a vital role in applications such as speech communication and transmission \cite{brandenburg1994iso, kroon1986regular, salami1994toll}. Recently, neural speech codecs have become key components for downstream generative tasks including speech language models \cite{borsos2023audiolm, wang2023neural, shen2023naturalspeech, junaturalspeech}, where codec-produced discrete codes serve as intermediate representations for speech generation.

The primary goal of speech codecs is to reconstruct high-quality speech at the lowest possible bitrate.
To achieve this, modern neural speech codecs typically adopt an encoder–quantizer–decoder architecture \cite{garbacea2019low, chen2021tenc, vali21_interspeech, lee2022progressive}. Among various vector quantization (VQ) \cite{van2017neural} techniques, residual vector quantization (RVQ) has become the de facto standard \cite{zeghidour2021soundstream}. RVQ employs a cascade of quantizers where each stage sequentially quantizes the residual errors from the previous one, and the final result is formed by summing the outputs of all stages. 
This hierarchical design greatly improves coding efficiency and has been adopted in numerous state-of-the-art codecs \cite{defossez2023high, wu2023audiodec, kumar2024high, ai2024apcodec, jiang2024mdctcodec, zheng2025ervq}.

Nevertheless, despite these advances, neural codecs are highly vulnerable in noisy acoustic environments \cite{niu2024ndvq}. Although they deliver excellent reconstruction quality on clean speech, even mild background noise can degrade their performance, reflecting limited robustness under realistic conditions \cite{casebeer2021enhancing}. This vulnerability arises because codecs are typically trained on clean speech, rendering them unprepared for noise-induced variability. As a result, their deployment in real-world scenarios is constrained by degraded perceptual quality under unavoidable background noise.

In this paper, we identify a key underlying cause of this vulnerability: the discrete and unstable nature of quantization decisions in RVQs. Each encoder feature is deterministically mapped to the nearest codeword in a codebook. However, even minor perturbations from environmental noise can push features across decision boundaries, resulting in entirely different codeword assignments and consequent reconstruction errors. A conventional remedy is to train codecs on paired noisy–clean speech data to promote robustness. Yet this approach suffers from two inherent drawbacks: (i) collecting sufficiently diverse noisy–clean pairs incurs additional data costs, and (ii) the learned robustness is often noise-specific, leading to poor generalization to unseen acoustic conditions and even potential degradation in clean-speech performance.

Motivated by these observations, we pose the central question: can we enhance noise robustness by addressing the instability of quantization decisions without relying on noisy training data? To this end, we propose a resource-efficient training strategy that simulates noise perturbations directly at the codeword level. Specifically, instead of deterministically selecting the nearest codeword, our approach employs probabilistic top-K sampling to mimic the stochastic variability induced by noise. To ensure stable optimization, we introduce a progressive training scheme that gradually applies probabilistic sampling from the last quantizer to the first in RVQ, incrementally enhancing robustness in a controlled manner. We validate the effectiveness of the proposed method on two advanced codecs, Encodec \cite{defossez2023high} and WavTokenizer \cite{ji2024wavtokenizer}. Experimental results show that our method significantly improves performance under noisy conditions while even enhancing coding quality on clean speech.

\section{Proposed Methods}
\label{sec:method}

\subsection{Motivation}

 \begin{figure}[tbp]
    \centering
    \includegraphics[width=0.95\linewidth]{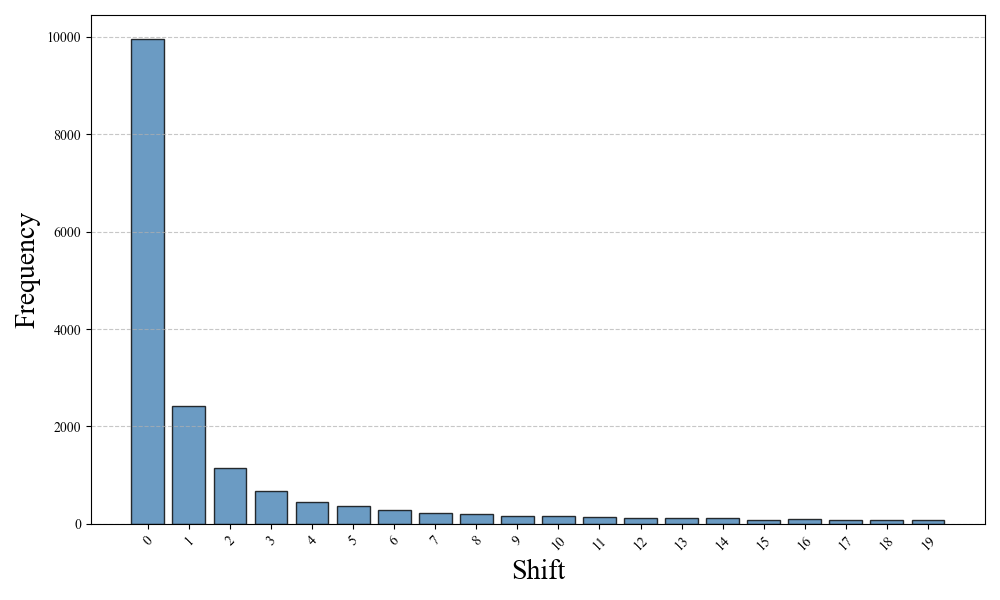} 
    \caption{
    Frequency distribution of codeword shifts observed in the first RVQ stage of Encodec \cite{defossez2023high} at 24 kHz sampling rate and 6 kbps bitrate. The test compares the quantized output for 120 clean speech signals from the VCTK dataset \cite{vctk} against their noisy counterparts generated by adding 15 dB SNR noise from the DEMAND dataset \cite{thiemann2013demand}. A shift of k-1 indicates that the codeword chosen for the noisy signal was the k-th closest candidate for the original clean signal. 
    }
    \label{fig:code_shift_histogram}
\end{figure}

Neural speech codecs are typically trained on clean speech, whereas real-world communication scenarios often involve background noise. As a result, codecs trained exclusively on clean data lack robustness and may degrade significantly even under mild noise conditions. 
To investigate this vulnerability, we analyze the quantized codes produced by the first quantizer (codebook size = 1024) in Encodec's \cite{defossez2023high} RVQ under noisy inputs.

Formally, given an encoding feature $\mathbf{z}_n$ to be quantized for noisy speech at the frame level, it is mapped to the nearest code vector $\mathbf{e}_{m_{n}}$ in the codebook based on Euclidean distance, with the corresponding codeword $m_{n}$ determined as
\begin{equation}
\label{Eq:closest distance}
    m_n =\arg \min_{j} ||\mathbf{z}_n-\mathbf{e}_j||_2,
\end{equation}
where $\mathbf{e}_j$ represents a code vector in the codebook, and the quantized result is $\hat{\mathbf{z}}_n=\mathbf{e}_{m_{n}}$. 
For the corresponding clean speech feature $\mathbf{z}_c$, from which $\mathbf{z}_n$ is derived by adding mild noise, the top-K closest codewords are given by:
\begin{equation}
\label{Eq:top-k}
\{ m_{c_1}, \ldots, m_{c_K} \} 
= \arg \min_{\substack{S \subseteq \mathcal{C} \\ |S| = K}}
\sum_{m \in S} \| \mathbf{z}_c - \mathbf{e}_m \|_2 ,
\end{equation}
where $\mathcal{C}$ denotes the codebook and $S$ is a subset containing $K$ code vectors, and the quantized result is $\hat{\mathbf{z}}_c=\mathbf{e}_{m_{c_{1}}}$. 
Our analysis reveals that the quantized code $m_n$ for  noisy speech often falls within the top-K candidates of its clean counterpart. The \textbf{codeword shift} is then defined as 
\begin{equation}
    \text{Shift}(m_n) = k-1 \quad \text{if } m_n = m_{c_{k}},
\end{equation}
which measures how far the noisy codeword deviates from the top-1 codeword of the corresponding clean feature.

As illustrated in Figure \ref{fig:code_shift_histogram}, while a zero shift (i.e., $m_n = m_{c_1}$) is the most frequent outcome, the pronounced long tail of non-zero shifts ($k>1$) reveals the fragility of the quantization process. This instability stems from the deterministic nearest-neighbor mapping, whose hard decision boundaries mean that even slight noise perturbations can push a feature vector into an adjacent Voronoi cell, thus altering the chosen codeword. These seemingly subtle shifts at the quantization stage accumulate across RVQ layers and manifest as audible artifacts in the reconstructed speech.
This observation motivates us to investigate whether noise robustness can be introduced directly at the quantization stage by simulating such perturbations, rather than relying on noisy–clean data pairs.


\subsection{Probabilistic Top-K Sampling}
Building on the above analysis, we propose a training strategy that enhances the robustness of neural speech codecs to unseen noise while requiring only clean data for training. 
Specifically, we replace the nearest-neighbor criterion for codeword selection during training with a probabilistic top-K sampling scheme in each VQ of the RVQ.

In the conventional VQ process, an encoding feature $\mathbf{z}$ is mapped to the nearest code vector $\mathbf{e}_m$ in the codebook using the closest Euclidean distance, where $m$ is defined in Equation \ref{Eq:closest distance}. 
In our proposed strategy, however, the quantization result is no longer restricted to the closest codeword. Instead, we first identify the top-K closest candidates $\{\mathbf{e}_{m_1},\dots,\mathbf{e}_{m_K}\}$ according to Equation \ref{Eq:top-k}, and then sample from them using a distance-based probability distribution. The probability distribution $\mathbf{P} = \{P_{m_1},\dots,P_{m_K}\}$ for top-K sampling is defined as:
\begin{equation}
    P_{m_i}=\frac{\exp{(-d_{m_i}/\tau)}}{\sum_{i=1}^K\exp{(-d_{m_i}/\tau)}},
\end{equation}
where $d_{m_i}=||\mathbf{z}-\mathbf{e}_{m_i}||_2$ and $\tau>0$ is a temperature hyperparameter controlling the sharpness of the distribution. 
The final quantized result $\hat{\mathbf{z}}$ is sampled from this distribution and used in the subsequent training process, while quantization reverts to the standard nearest-neighbor rule at inference.

A key design choice is the use of distance-based probabilistic top-K sampling rather than uniform random sampling. This decision is motivated by the above observation that mild noise perturbations are more likely to push a feature vector across a nearby decision boundary than into a distant region of the codebook. Accordingly, our probabilistic sampling strategy explicitly models this locality of perturbations by assigning higher probabilities to codewords closer to the clean feature. This contrasts with uniform random sampling, which disregards geometric structure and may generate unrealistic perturbations. 
By incorporating distance-based probabilistic top-K sampling, our proposed method effectively exposes the decoder to plausible variations at the codeword level during training, thereby improving its resilience to genuine noise at test time. This approach is resource-efficient and enhances the codec's robustness to noise without requiring paired noisy and clean data.

\subsection{Progressive Training Strategy}
The residual nature of RVQ provides a natural hierarchy in which different quantizers capture information at different scales: the first VQ encodes the most salient structural components of the signal, while subsequent VQs progressively refine the representation by encoding residual details, with the final $N$-th VQ modeling the finest acoustic nuances. 
Leveraging this hierarchy, we design a progressive training strategy that can be viewed as a form of curriculum learning.
Specifically, for an RVQ containing $N$ VQs, we begin by first applying probabilistic top-K sampling only to the final $N$-th VQ, where perturbations influence fine-grained residual details and thus exert minimal impact on overall reconstruction quality. At this stage, all preceding VQs still use the standard nearest-neighbor deterministic rule, ensuring stability. After the model adapts, we shift the perturbation one stage earlier, applying probabilistic sampling to the $(N-1)$-th VQ while keeping all other VQs deterministic. This process is repeated iteratively, moving layer by layer from the last VQ toward the first. In this way, perturbations are introduced in a controlled, curriculum-like manner, gradually extending from peripheral residual details to the core structural components of the signal representation.
This easy-to-hard learning schedule mitigates the risk of destabilizing training by prematurely perturbing the primary structural features of the signal. Instead, robustness is built up gradually and consistently across RVQ layers, enabling the codec to internalize noise resilience in a stable and effective manner.

The overall progressive probabilistic top-K sampling training strategy is summarized in Algorithm \ref{alg:topk_sampling}.

\begin{algorithm}[t]
\small
\caption{Progressive Probabilistic Top-K Sampling in RVQ}
\label{alg:topk_sampling}
\begin{algorithmic}[1]
\REQUIRE Encoding feature $\mathbf{z}$,  $N$ VQ codebooks $\{\mathbf{e}^n_j\}_{n=1}^N$, number of top candidates $K$
\ENSURE Quantized result $\hat{\mathbf{z}}$

\FOR{$l = N$ to $1$}
    \STATE $\hat{\mathbf{z}} = 0$, $\textbf{residual} = \mathbf{z}$
    \FOR{$n = 1$ to $N$}
        \STATE Compute distances $d_j = ||\textbf{residual} - \mathbf{e}_j^n||_2$
        \IF{$n == l$}
            \STATE Select top-$K$: \\ $\{ m_{c_1}, \ldots, m_{c_K} \} = \operatorname{Top-K}_j \| \mathbf{z}_c - \mathbf{e}_j \|_2$
            \STATE Compute probabilities: $P_{m_i} = \frac{\exp(-d_{m_i}/\tau)}{\sum_{j=1}^K \exp(-d_{m_j}/\tau)}$
            \STATE Sample $m \sim \text{Categorical}(P_{m_1}, \dots, P_{m_K})$
        \ELSE
            \STATE Select closest: $m = \arg \min_{j} d_j$
        \ENDIF
        \STATE $\hat{\mathbf{z}} = \hat{\mathbf{z}} + \mathbf{e}_m^n$, $\textbf{residual} = \textbf{residual} - \mathbf{e}_m^n$
    \ENDFOR
    \STATE Update gradients for components after $l$-th VQ.
\ENDFOR
\end{algorithmic}
\end{algorithm}

\section{Experiments}
\label{sec: Experiments}

\subsection{Datasets}

For our experiments, we used a subset of the VCTK-0.92 speech corpus \cite{vctk}, following the setup in \cite{ai2024apcodec}. 
To simulate mild background noise in real-world speech communication scenarios, noisy test samples were generated by mixing clean speech with noise. Specifically, 18 types of noise from the DEMAND database \cite{thiemann2013demand} were randomly added to the speech at SNRs of 15 and 10 dB. Voice activity detection (VAD) was applied to the dataset to extract non-silent segments for training and testing,
ensuring that the model focused on the most relevant portions of the speech signal.

\subsection{Implementation Details}
We employed Encodec \cite{defossez2023high}, an advanced streamable neural speech codecs, for our experiments. The model was configured with a 24 kHz sampling rate and a 6 kbps bitrate, comprising six VQs, each with a codebook size of 1024. 
We also conducted experiments on WavTokenizer \cite{ji2024wavtokenizer}, a state-of-the-art single-codebook neural speech codec. This model was configured with a 24 kHz sampling rate and a 0.9 kbps bitrate, using a single VQ with a codebook size of 4096. Both models were first trained on the training set using the traditional closest-distance strategy until convergence, with a learning rate of 3e-4. 
Subsequently, we fine-tuned the pre-trained Encodec model using the proposed progressive probabilistic top-K sampling strategy, and the WavTokenizer model using the standard probabilistic top-K sampling strategy, with a learning rate of 1e-4.
The choice of $K$ was guided by our preliminary analysis of the codeword shift phenomenon, which revealed that most shifts occur within the 10 nearest candidates of the corresponding clean features. Accordingly, we set $K=10$ to balance realistic perturbations modelling and computational efficiency. For the softmax temperature $\tau$, we observed that a large $\tau$ (e.g., 10) flattens the distribution toward random sampling, while a small $\tau$ (e.g., 1) sharpens it excessively, collapsing to nearest codeword selection. We therefore adopt a moderate value of $\tau=5$ to achieve a trade-off between diversity and stability. For experimental efficiency, all analysis experiments were conducted on Encodec.

\vspace{-1mm}
\subsection{Evaluation Metrics}
Several evaluation metrics were employed to assess the performance of our proposed methods.
These metrics include Scale-Invariant Signal-to-Distortion Ratio (SI-SDR) for energy-based speech quality assessment, Perceptual Evaluation of Speech Quality (PESQ) \cite{rix2001perceptual} for speech quality assessment based on perceptual models ranging from –0.5 to 4.5, and Short-Time Objective Intelligibility (STOI) \cite{taal2010short} for measuring speech intelligibility on a scale from 0 to 1. Additionally, UTMOS\footnote{\url{https://github.com/sarulab-speech/UTMOS22}.} \cite{saeki22c_interspeech}, a non-intrusive pre-trained scoring network, was adopted to evaluate the naturalness of the decoded speech in the range of 1 to 5. These metrics offer a comprehensive assessment of the noise robustness of Encodec trained using different strategies.

\begin{table}[tbp]
  \centering
  \caption{Evaluation metrics for noisy and clean speech coding with and without the proposed training strategy. The best results are highlighted in \textbf{bold}. Values in (parentheses) represent p-values from paired t-tests, where p-value $<$ 0.05 indicates statistical significance.}
    \label{Table:Overall Performance}
    \resizebox{\linewidth}{!}{%
    \begin{tabular}{c|c|c|cccc}
    \toprule
    Model & Noise SNR & Proposed & SI-SDR & PESQ  & STOI  & UTMOS \\ \hline
    \multirow{9}[4]{*}{Encodec} & \multirow{3}[2]{*}{15dB} & No    & 4.519 & 2.399 & 0.935 & 3.475 \\
          &       & Yes   & \textbf{5.232} & \textbf{2.466} & \textbf{0.938} & \textbf{3.586} \\
          &       &       & (4.45e-157) & (1.89e-3) & (4.76e-5) & (7.76e-77) \\
\cline{2-7}          & \multirow{3}[2]{*}{10dB} & No    & 3.896 & 2.021 & 0.915 & 3.243 \\
          &       & Yes   & \textbf{4.524} & \textbf{2.068} & \textbf{0.918} & \textbf{3.352} \\
          &       &       & (1.80e-12) & (3.89e-2) & (3.19e-2) & (8.58e-10) \\ 
\cline{2-7}          & \multirow{3}[2]{*}{Clean} & No    & 4.794 & 3.410 & 0.964 & 3.732 \\
          &       & Yes   & \textbf{5.513} & \textbf{3.552} & \textbf{0.967} & \textbf{3.854} \\
          &       &       & (7.81e-26) & (1.78e-32) & (2.74e-11) & (1.12e-54) \\ \hline
    \multirow{9}[4]{*}{WavTokenizer} & \multirow{3}[2]{*}{15dB} & No    & 0.973 & 1.792 & 0.875 & 3.607 \\
          &       & Yes   & \textbf{1.222} & \textbf{1.839} & \textbf{0.878} & \textbf{3.654} \\
          &       &       & (9.09e-4) & (5.37e-9) & (4.11e-2) & (9.10e-7) \\
\cline{2-7}          & \multirow{3}[2]{*}{10dB} & No    & 0.511 & 1.633 & 0.855 & 3.486 \\
          &       & Yes   & \textbf{0.791} & \textbf{1.685} & \textbf{0.860} & \textbf{3.542} \\
          &       &       & (8.73e-4) & (8.51e-6) & (4.92e-2) & (2.85e-6) \\
\cline{2-7}          & \multirow{3}[2]{*}{Clean} & No    & 1.191 & 2.068 & \textbf{0.897} & 3.884 \\
          &       & Yes   & \textbf{1.217} & \textbf{2.122} & 0.896 & \textbf{3.946} \\
          &       &       & (7.31e-3) & (4.84e-11) & (4.40e-1) & (3.34e-17) \\
    \bottomrule
    \end{tabular}%
    }
  \label{tab:addlabel}%
\end{table}%

\begin{table}[t]
  \centering
  \caption{Evaluation metrics for clean and noisy speech coding using Encodec with different training strategies. The best results are highlighted in \textbf{bold}. \underline{Underline} characters indicate the sub-optimal results.}
  \label{Table:Analysis}
    \resizebox{\linewidth}{!}{%
    \begin{tabular}{c|c|cccc}
    \toprule
    Speech & Strategy & SI-SDR & PESQ  & STOI  & UTMOS \\ \hline
    \multirow{3}[2]{*}{Noisy} & Proposed & \textbf{5.232} & 2.466 & 0.938 & 3.586 \\
          & Closest* & 4.715 & \textbf{2.970} & \textbf{0.943} & \textbf{3.769} \\
          & Proposed\dag & 4.762 & 2.426 & 0.936 & 3.539 \\ \hline
    \multirow{3}[2]{*}{Clean} & Proposed & \textbf{5.513} & \textbf{3.552} & \textbf{0.967} & \textbf{3.854} \\
          & Closest* & 4.731 & 3.179 & 0.954 & 3.793 \\
          & Proposed\dag & 5.063 & 3.455 & 0.965 & 3.798 \\
    \bottomrule
    \end{tabular}%
    }
\end{table}%

\subsection{Overall Performance}
Table \ref{Table:Overall Performance} provides a comprehensive comparison of the performance of our method and the baselines on both clean and noisy speech\footnote{Speech samples are available at: \url{https://zhengrachel.github.io/NoiseRobustAudioCodec/}.}. 
For noisy speech coding, our proposed method yields consistent improvements across all evaluation metrics compared with the standard RVQ baseline, e.g., UTMOS improves from 3.475 to 3.586 at 15 dB SNR for Encodec, indicating a perceptible gain in the naturalness of decoded speech. Moreover, it also improves the decoding quality for clean speech in most metrics, likely due to the perturbations introduced at the codeword level, which act as a form of data augmentation mechanism.

\begin{table}[]
\centering
\caption{Evaluation metrics for noisy speech coding by the Encodec trained with the proposed method and the baseline method under three different noise types. Best results are highlighted in \textbf{bold}.}
\label{Table:Performance on 3 Noise Types}
\resizebox{\linewidth}{!}{%
\begin{tabular}{c|c|cccc}
\toprule
Noise                      & Strategy & SI-SDR         & PESQ           & STOI           & UTMOS          \\ \hline
\multirow{2}{*}{DWASHING} & Proposed & \textbf{5.167} & \textbf{3.372} & \textbf{0.958} & \textbf{3.811} \\
                        & Closet*  & 4.829          & 3.154          & 0.947          & 3.751          \\
                           \hline
\multirow{2}{*}{OOFFICE}  
                          & Proposed & \textbf{5.349} & \textbf{3.186} & \textbf{0.959} & \textbf{3.825} \\ 
                          & Closet*  & 4.628          & 3.181          & 0.951          & 3.800          \\ \hline
\multirow{2}{*}{TCAR}     
                          & Proposed & \textbf{5.239} & \textbf{3.254} & \textbf{0.956} & \textbf{3.821}  \\
                          & Closet*  & 4.736          & 3.130          & 0.944          & 3.751          \\
\toprule
\end{tabular}
}
\end{table}

\subsection{Analysis and Discussion}

\subsubsection{Comparison with Noise-Exposed Fine-Tuning}
To comprehensively evaluate our method, we compared it with a strong baseline where Encodec was fine-tuned end-to-end on paired noisy–clean data, denoted as \textit{Closest*}. The training noisy speech was generated by randomly mixing all 18 types of noise from the DEMAND database with the clean training set. It is important to note that \textit{Closest*} benefits from direct exposure to the same noise sources used in testing, whereas our approach was trained exclusively on clean speech and thus encounters test noise in a \textbf{zero-shot} manner.  

As shown in Table \ref{Table:Analysis}, \textit{Closest*} achieves higher scores than our method on certain noisy-speech metrics (e.g., PESQ). This outcome is expected: \textit{Closest*} is effectively coached on the specific test noise and therefore represents an \textbf{upper-bound reference} under matched conditions. However, this advantage comes with significant trade-offs. First, \textit{Closest*} degrades performance on clean speech, which is problematic for many real-world scenarios where speech often transitions between noisy and clean conditions. Second, its robustness is largely noise-specific, limiting generalization to unseen acoustic environments.
To further illustrate this distinction, Table \ref{Table:Performance on 3 Noise Types} compares the two methods on three representative noise types. Importantly, these noise conditions were present in the training set of \textit{Closest*}, but not in ours. Even under this disadvantage, our approach outperforms \textit{Closest*} on all three cases, providing strong evidence that the proposed perturbation-based training instills robustness to perturbations in general, rather than to particular noise types. This generalization advantage is critical for deployment in diverse and umpredictable real-world acoustic environments.

Overall, unlike \textit{Closest*}, which relies on noise-specific training, our method provides a more practical solution: it is data-efficient, preserves clean-speech quality, and generalizes to unseen noise, making it cost-effective and applicable in real-world environments.

\subsubsection{Effectiveness of Probabilistic Top-K Sampling}
\label{Section: probablistic top-k sampling}
Table \ref{Table:Analysis} also includes an ablation study comparing the proposed strategy to a variant where probabilistic sampling was replaced with random sampling (\textit{Proposed\dag}). Across both clean and noisy speech conditions, the probabilistic sampling method consistently outperforms random sampling across all metrics. 
This confirms that the effectiveness of our method lies not merely in introducing stochasticity, but in the way stochasticity is structured. 
By leveraging distance-based probabilities, the probabilistic sampling approach introduces a structured and effective perturbation during training. In contrast, uniform random sampling disregards geometric proximity and often generates implausible perturbations, leading to weaker robustness.

 \begin{figure}[tbp]
    \centering
    \includegraphics[width=\linewidth]{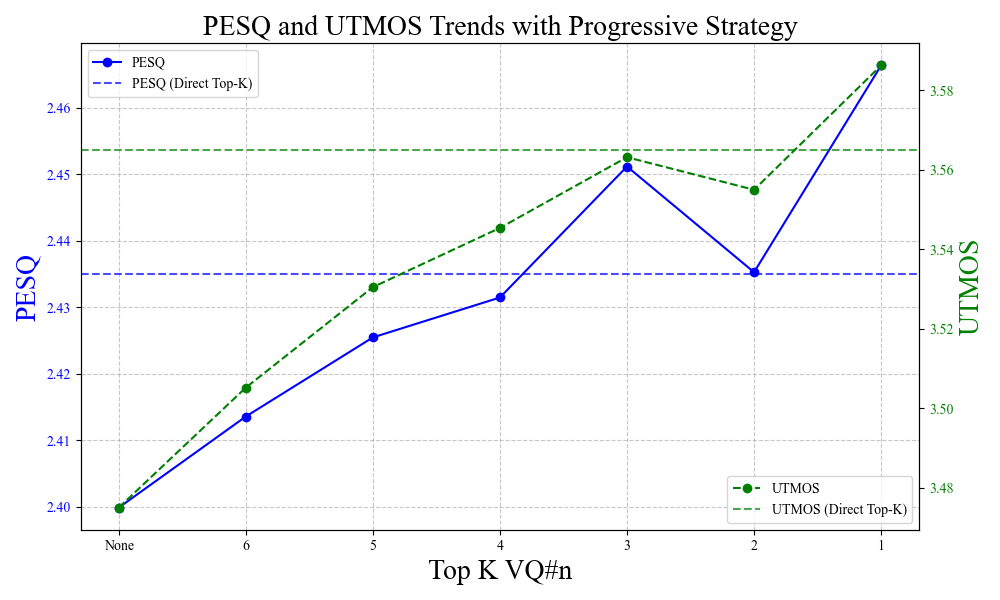} 
    \caption{PESQ and UTMOS trends during the progressive application of the probabilistic top-K sampling strategy from the 6th VQ to the 1st VQ. The left vertical axis represents PESQ scores, while the right vertical axis represents UTMOS scores.}
    \label{fig:pesq_utmos_variations}
\end{figure}

\subsubsection{Effectiveness of Progressive Training Strategy}

Figure \ref{fig:pesq_utmos_variations} illustrates the evolution of PESQ and UTMOS as probabilistic top-K sampling is progressively applied from the 6th to the 1st VQ in Encodec's RVQ. 
Both metrics exhibit a steady upward trend, indicating consistent improvements in perceptual quality and naturalness as robustness is gradually injected.
In contrast, applying probabilistic top-K sampling directly to all VQs (\textit{Direct Top-K}) yields inferior results, as early perturbation of core features destabilizes training. These findings confirm that the progressive strategy provides a controlled, curriculum-like schedule: starting with weaker perturbations at later VQs and progressively extending to earlier, more influential ones, thereby enabling the model to adapt smoothly to varying noise levels and develop robust representations.

\section{Conclusion}
In this paper, we proposed a resource-efficient training framework to enhance the noise robustness of neural speech codecs. By introducing perturbations at the codeword level through probabilistic top-K sampling and incorporating them via a progressive, curriculum-like schedule, our method directly addresses the instability of quantization decisions without requiring any noisy–clean data.
Extensive experiments on Encodec and WavTokenizer demonstrated that the proposed strategy not only improves robustness under noisy conditions but also preserves—and in some cases even enhances—performance on clean speech. 
Overall, this work highlights codeword-level perturbation modeling as an effective principle for building noise-robust speech codecs. Future directions include extending the framework to more streamable architectures and exploring its integration with large speech–language models.

\bibliographystyle{IEEEbib}
\bibliography{strings,refs}

\end{document}